# Responding to the AIDS epidemic in Angola

Brian G. Williams

South African Centre for Epidemiological Modelling and Analysis (SACEMA), Stellenbosch, South Africa
Correspondence to BrianGerardWilliams@gmail.com

**Summary**

The epidemic of HIV in Angola started later and stabilized at lower levels than in most countries in southern Africa. Some studies suggest that certain high risk groups account for a substantial part of the total number of infections. With a relatively small population and a relatively high gross domestic product, Angola is in a good position to intervene decisively to control HIV. The effectiveness, availability and affordability of potent anti-retroviral therapy (ART) make it possible to contemplate ending the epidemic of HIV/AIDS in Angola. In this paper we consider what would have happened without ART, the *No ART* counterfactual, the impact on the epidemic if the current roll-out of ART is maintained, the '*Current Programme*', the impact if coverage is rapidly increased to reach 90% of people with CD4$^+$ cell counts below 350/μL by 2015 and HIV-positive pregnant women are all offered ART for life ('*Option B+*'), the '*Accelerated Programme*', and what might be possible under the 2013 guidelines from the World Health Organization, starting in 2015 and reaching full coverage of ART by 2018, the *Expanded Programme*.

*Current Programme*

The number of people receiving ART in Angola has doubled in the last 3 years from 20 thousand in 2009 to 42 thousand in 2012. Compared to the *No ART* counter-factual the *Current Programme* will, by 2015, reduce the prevalence of HIV among adults *not* on ART from 2.5% to 1.8%, the annual incidence from 0.24% to 0.18%, and AIDS related deaths from 0.20% to 0.10% p.a. By 2015 the *Current Programme* will have averted 24 thousand new infections in adults, 8.6 thousand new infections in new born children, saved 51 thousand lives, and US$51M.

*Accelerated Programme*

Under the *Accelerated Programme* the number of people starting treatment would have to be doubled between now and 2015. Compared to the *Current Programme* this will further reduce the prevalence among adults *not* on ART from 1.8% under the to 1.4%, the incidence from 0.18% *p.a.* to 0.14% *p.a.*, and AIDS related deaths from 0.10% *p.a.* to 0.05%. By 2015 the *Accelerated Programme* will have averted an additional 4 thousand new infections in adults, 1.7 thousand infections in new born children, saved 7 thousand lives, and US$51M.

*Expanded Programme*

Under the *Expanded Programme* 90% of all HIV positive people will have access to immediate treatment by 2020. Compared to the *Accelerated Programme* this will further reduce the prevalence among adults *not* on ART in 2020 from 0.84% to 0.14%, incidence from 0.11% *p.a.* to 0.02% *p.a.*, and AIDS deaths from 0.01% *p.a.* to elimination under the *Expanded Programme*. By 2020 the *Accelerated Programme* will have averted an additional 27 thousand new infections in adults, 6 thousand infections in new born children, saved an additional 4 thousand lives, and US$51M.

    If Angola is to reach the 2015 targets in the *Presidents Acceleration Plan*, several key areas must be addressed. Testing services will need to be expanded rapidly supported by a mass testing campaign, so as to diagnose people with HIV and enrol them in treatment and care. A regular and uninterrupted supply of drugs will have to be assured. The quantity and quality of existing health staff will need to be strengthened. Community health workers will need to be mobilized and trained to encourage people to be tested and accept treatment, to monitor progress and to support people on treatment; this in turn will help to reduce stigma and discrimination, loss to follow up of people diagnosed with HIV, and improve adherence for those on treatment. Effective monitoring and evaluation systems will have to be in place and data collection will have to be extended and improved to support the development of reliable estimates of the current and future state of the epidemic, the success of the programme, levels of viral load suppression for those on ART and the incidence of infection.

    If Angola succeeds in reaching the targets in the Presidents *Acceleration Plan*, this would provide the foundation for development and implementation of the *Expanded Programme*, ensuring that almost all HIV-positive people have access to ART. This will require an even greater initial commitment, political as well as financial, but should lead to the elimination of transmission and, eventually, the end of the pandemic.

## Introduction

The prevalence of HIV infection among women attending ante-natal clinics (ANC) in Angola has now stabilized at a prevalence that is lower than in most of southern Africa. The *Current Programme* of anti-retroviral therapy (ART) was started in 2003 and has already reduced AIDS related mortality significantly. Following the current recommendations of the World Health Organization (WHO),[1] the Angolan Government is considering expanding the provision of ART. Here we estimate the impact on the epidemic if 1) the *Current Programme* of ART provision is continued, 2) the *Accelerated Programme* is successfully implemented so that 90% of those currently in need of treatment have access to treatment by 2015, 3) the provision of ART is further expanded after 2015 in accordance with the new (2013) WHO guidelines for starting treatment.[1]

    We start from data on the prevalence of HIV among

women attending ante-natal clinics in ANC facilities in the districts of Angola (Appendix 1). We use these data to estimate the year in which the HIV epidemic started and the asymptotic prevalence in each district. We combine these estimates to obtain a national estimate of the time trend in the prevalence of HIV. We fit the national trend in prevalence to a dynamical model to estimate current and to project future, trends in prevalence, incidence, treatment needs and deaths. We estimate the *per capita* cost of HIV/AIDS, including the cost of providing drugs, providing support to people on ART, and hospitalization and access to primary health care facilities for people who are not on ART and develop AIDS related conditions. We do not include the social and economic costs incurred when people die of AIDS so that this calculation is conservative with regards to the overall cost to society.

## Data

The ANC data (Appendix 1) are sparse; in some years no data were collected and most facilities reported for only a few years. The timing of the epidemic is important: the earlier the epidemic started the more people are likely to have advanced HIV-disease and *vice versa*. We therefore consider each data set in separately and then combine them to arrive at an epidemic curve for the country as a whole. Further details are given in Appendix 1.

In generalized epidemics the prevalence of HIV is higher in pregnant women than in women in the general population and is higher in women than in men. The prevalence of infection among ANC women has therefore been scaled down by 20% to give the prevalence in all adults following UNAIDS recommendations.[2]

Table 1. Annual cost of days in hospital, primary health care visits and counselling and testing from Granich *et al.*[3] Costs are for the year 2013. The number of inpatient days and primary health care visits are based on data for South Africa.[4]

| Item | Cost (US$) |
|---|---|
| Inpatient days not on ART | 568 |
| Inpatient days on ART | 138 |
| Primary health care visits not on ART | 154 |
| Primary health care visits on ART | 269 |
| Counselling and testing per test | 20 |
| Community care and support on ART | 250 |

Costs include hospitalization, primary health care and treatment.[3] Drug costs are in Table 2. We do not discount future costs but the data presented here could be used to discount costs and benefits at any desired rate.

Table 2. Annual cost of ART in South Africa.

| Year | Cost (US$) |
|---|---|
| 1995 | 10,000[5] |
| 2003 | 1,700[6] |
| 2006 | 730[7] |
| 2009 | 188[3] |
| 2014 | 100[8,9] |

The cost of combination ART in South Africa has declined dramatically since it first became available (Table 2 and Figure 1). Since drug costs are likely to decline in Angola in the near future we set all drug costs to those for South Africa[4] (red line, Figure 1).

## Epidemic Model

The model is a standard dynamical model discussed in detail elsewhere.[10,11] It includes uninfected people who are susceptible to infection while infected people go through four stages of infection to death to match the known Weibull survival for people infected with HIV but not ART.[12] To account for heterogeneity in the risk of infection we let the transmission parameter decline with the prevalence of HIV[10,11] according to a cumulative Weibull distribution with a shape parameter of 2.25 (Appendix 2).

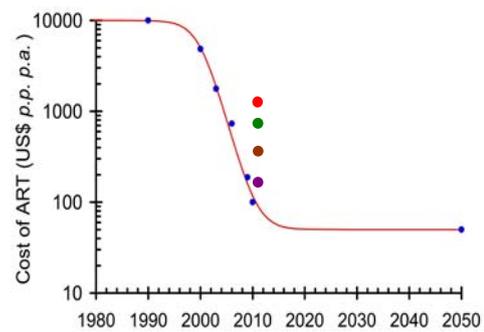

Figure 1. Annual cost of ART per person per year in South Africa[4] (blue dots; Table 1) fitted to a logistic curve (red line). Large dots give current prices in Angola: purple: children; brown: first line; green: PMTCT; red: second line (Appendix 4).

## Fitting the model

We first fit the model to the prevalence data without including ART to get a null model against which to compare the impact of ART. We vary the prevalence of infection in 1980, which determines the timing of the epidemic, the rate of increase and the rate at which the risk of infection declines as the prevalence increases as this, in turn, determines the peak value of the prevalence. We also vary the parameter that determines the shape of the function relating transmission to prevalence (Appendix 2). The fitted values are in Table 3.

To model the provision of ART in the *Current Programme* we assume that certain proportions of people in the third and fourth stages of HIV infection start ART and that coverage increases logistically. We then vary the rate and timing of the increase and the proportion of people starting treatment in each stage to match the reported ART coverage.[2] To model the *Accelerated Programme* we further increase the proportion of people starting treatment in WHO clinical stages 3 and 4 to 80% and set the rate at which this roll-out happens to 2.0/year. Under the *Accelerated Programme* 90% of all pregnant women are started on ART for life. To model the *Extended Programme* we include active case finding and assume that coverage increases logistically at a realistic rate, timing and asymptotic coverage. We assume that all those that are found to be HIV-positive, in any stage of infection, are eligible for treatment. The parameter values are in Table 4.

Table 3. The birth, background mortality and transition rates are fixed;[10] the other parameters are varied to optimize the fit to the trend in the prevalence of HIV shown in Figure 2 (below).

| Parameter | Value |
|---|---|
| Adult population in 2013 (millions) | 9.27 |
| Population growth rate (percent/yr) | 2.2 |
| Background mortality (percent/yr) | 1.8 |
| Force of infection/yr | 0.726 |
| Prevalence at which transmission is halved (%) | 1.3 |
| Shape parameter (see Appendix 2) | 2.25 |
| Transition rate/yr between HIV stages off ART | 0.348 |
| Transition rate/yr between HIV stages on ART | 0.087 |

Table 4. Model parameters for logistic functions. *Coverage* gives the asymptotic coverage; *Rate* the exponential rate of increase; *Half-max* the year when coverage reaches half the maximum value. *OR* is the odds-ratio for the number tested to the prevalence of HIV in that stage.

| | Parameter | Value |
|---|---|---|
| Passive case finding: Stage 4 | Coverage | 0.19 |
| | Rate *per annum* | 1.00 |
| | Half-max. (year) | 2000.0 |
| | OR (testing) | 1 |
| Passive case Finding: Stage 3 | Coverage | 0.10 |
| | Rate *per annum* | 1.00 |
| | Half-max. (year) | 2000.0 |
| | OR (testing) | 2 |
| Active case finding: testing | Coverage | 0.90 |
| | Rate *per annum* | 2.00 |
| | Half-max. (year) | 2016.0 |
| | Test interval (yrs) | 1.00 |
| | OR (testing) | 10 |
| Active case finding: take-up | Coverage | 0.90 |
| | Rate *per annum* | 2.00 |
| | Half-max. (year) | 2016.0 |

**Testing rates and costs**

In the model a proportion of HIV-positive people are tested a certain number of times each year and a proportion of those that test positive start ART. However, many people who are not infected will also be tested and in order to determine the cost of testing we need to know the case detection rate, that is the proportion of all those tested that are infected with HIV. Under passive case-finding, people present to a health-service in Stages 3 or 4 of HIV infection. We assume that all those that present in Stage 4 will be infected with HIV but that only 50% of those that present in Stage 3 will be infected with HIV.

Under active case finding we assume that only 2% of those that are tested will be infected with HIV. Since we also have to avoid the mathematical possibility that we test more people than there are in the population we set the odds ratio for the number tested to the number that are HIV-positive to 1 for stage 4, 2 for stage 3, and 50 for active case-finding; when the prevalence is low the number that are positive is 100%, 50% or 2%, respectively, of the number tested but when the prevalence is high the number tested in each test interval never exceeds the total population.

**Results**

The fitted data and implied trends in prevalence, incidence and mortality as well as the number on treatment and the rate at which people start treatment, are shown in Figure 2 using the parameter values in Table 3 and Table 4. Figure 2 gives, from left to right, the prevalence of infection, the incidence of infection and of treatment, and the costs. From top to bottom the graphs give the counterfactual under the *No ART* counterfactual, the *Current Programme*, the *Accelerated Programme*, and the *Expanded Programme*.

*No ART*

Figure 2A gives the prevalence of HIV, Figure 2B the implied incidence (red line) and mortality (black line) and Figure 2C the implied costs of in-patient and out-patient care[3] (red line). Prevalence rises rapidly to a steady state and remains fairly constant at about 2.5% after 2005. Incidence peaks, declines as the epidemic saturates and people start to die, and levels off at about 0.25% per year. Mortality rises about ten years after the incidence, reflecting the mean life-expectancy of people with HIV, and levels off at 0.19% per year. The annual cost to the health system would have increased to US$8 per adult per year, or US$72M per year, in 2013 and then remained constant. The rapid decline in incidence from 0.57% *p.a.* in 2005 to 0.22% in 2010 (Figure 2B red line) reflects the intrinsic dynamics of the epidemic through the saturation of the relatively small number of Angolan's that are at risk of infection and the increasing mortality as those that are infected begin to die.

*Current Programme*

The provision of ART on a significant scale in the public sector in Angola began in about 2005. Following the then recommendations of the World Health Organization (WHO)[13] infected people only started ART when their $CD4^+$ cell count fell to 200/μL or they were in Clinical Stages 3 or 4.[14] In 2006 WHO increased the $CD4^+$ cell count at which people should start ART to 350/μL and the Angolan government followed suit. Figure 2D gives the HIV prevalence of those not on ART (red line), the reported number (green dots) the fitted number (green line) of people on ART and all infected people (blue line); Figure 2E gives the incidence of HIV (red line), the rate at which people start ART (green line), the mortality of those on ART (black line) and of those not on ART (grey line). Figure 2F gives the cost of ART (green line)[3] and the cost of ART plus the cost of in-patient and out-patient care (red line).[3]

By 2015 the prevalence of HIV in adults is expected to be 2.5% (Figure 2D: blue line), the prevalence of HIV in adults *not* on ART (Figure 2D: red line) 1.8%, and the prevalence of adults on ART 0.63%. This means that about 25% of all HIV positive adults will be receiving ART or about 50% of adults with a $CD4^+$ cell count below 350/μL.

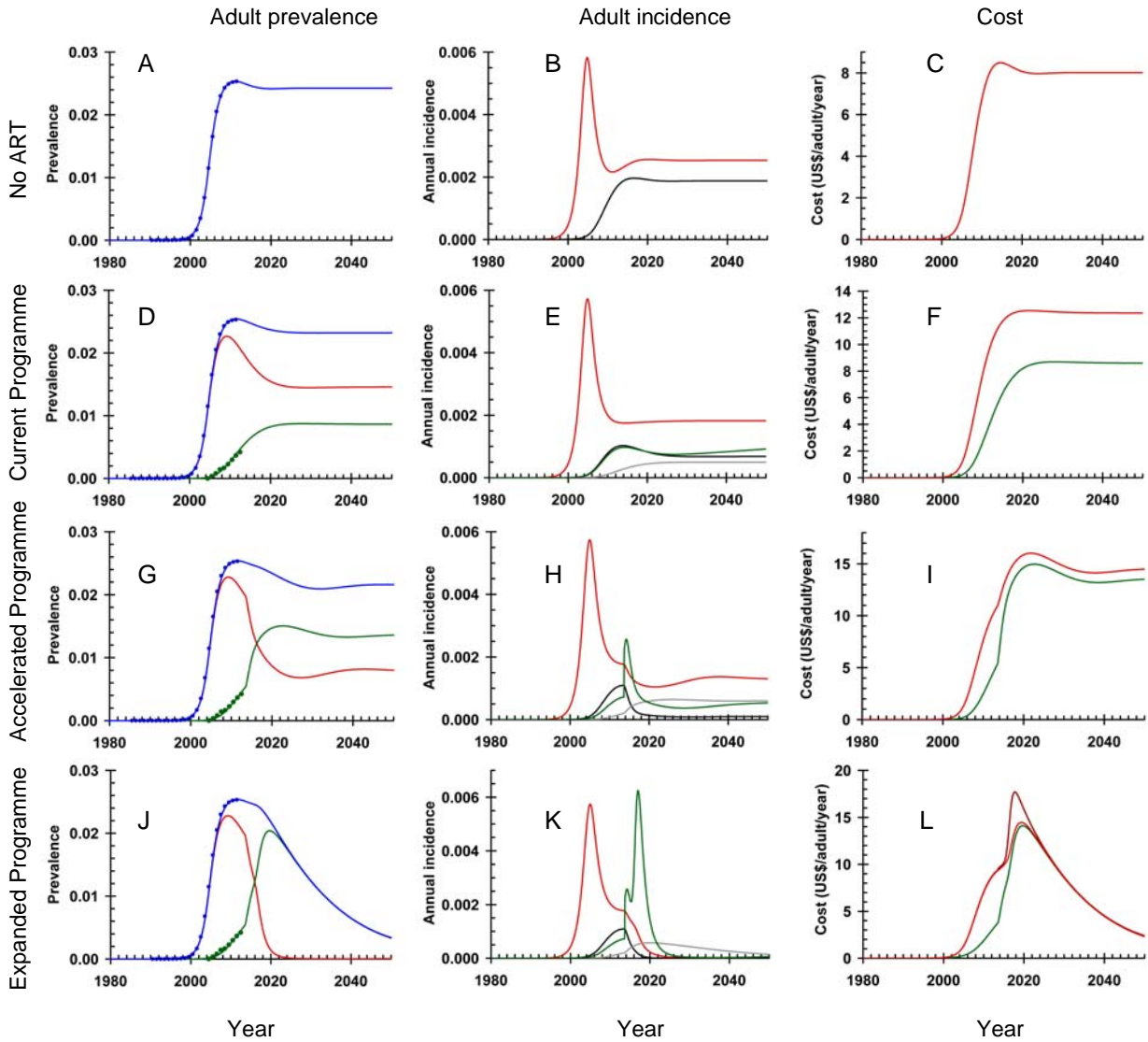

Figure 2. Top row*: No ART*. Second row: *Current Programme*. Third row: *Accelerated Programme*. Bottom row: *Expanded Programme*. Left*:* HIV prevalence. Blue: Data (with 95% confidence limits) and fitted line. Red: not on ART. Green: Data for those on ART and fitted line. Middle*:* HIV incidence. Red: Incidence of HIV. Green: incidence of treatment. Black: mortality not on ART. Grey: mortality on ART. Right: Cost. Green: ART. Red: Green plus hospitalization. Brown: Red plus cost of testing.

*Accelerated Programme*

The *Accelerated Programme* assumes that the number of people on treatment will be doubled by the end of 2015. (Figure 2G). This will require a massive campaign of testing and treatment with the associated logistical problems.

By 2015 the incidence will have fallen further from 2.4% *p.a.* without ART to 1.8% *per annum* under the *Accelerated Programme* (Figure 2E: red line). The rate at which new people start ART will peak at 0.09% *per annum* in 2014 (Figure 2E: green line) but will fall after that. AIDS related deaths among adults will have fallen from about 0.2% *p.a.* without ART to 0.1% *per annum* under the *Accelerated Programme* (Figure 2E: black line). Without ART about 107 thousand people would have died by 2015; the provision of ART has reduced this to 71 thousand saving 36 thousand lives. People will continue to die on ART (Figure 2E: grey line) but mainly from natural causes other than HIV.

Both the *Current Programme* and the *Accelerated Programme* have saved and will continue to save many lives but their impact on incidence and prevalence is much less; providing ART relatively late in the course of infection keeps people alive but only after they have infected several other people. There will also be a small but significant cost saving of about US$32M out of a total cost of up to 2015 of about US$547M. The total cost includes care and support, (Figure 2F: green line) and in-patient and out-patient care (Figure 2F: difference between red and green lines).

*Expanded Programme*

The impact of the *Accelerated Programme* will be greater on mortality than on incidence or prevalence and the savings that accrue encourage one to consider the impact of expanding the programme further. If the Angolan government chooses to adopt the 2013 guidelines of the World Health Organization[15] about 90% of all HIV-positive people will be eligible for treatment as soon as they are found to be HIV-positive.[16] The current (2013) guidelines of the International AIDS Society[17] and the Department of Health and Human Services (DHHS)[18] both recommend treatment for those infected with HIV without regard to their $CD4^+$ cell count on the grounds that this is in the best interests of the individuals concerned and has the added benefit of reducing the likelihood that they will infect their partners.

We therefore consider what would happen under a policy of universal access in which 90% of all adults, not on ART, are tested each year and 90% of those that test positive are started on ART starting in 2015 and reaching full coverage by 2020.

Universal access to early treatment should eliminate HIV transmission (Figure 2J and Figure 2K: red lines) and AIDS related deaths by 2022 (Figure 2K: black line) but there will still be a very large number of HIV positive people on ART (Figure 2J: green line) who will have to be maintained on treatment for the rest of their lives. The rate at which people need to be started on treatment will increase to about ten times the present rate (Figure 2K: green line) in 2015 but after that will fall rapidly as transmission and the generation of new cases fell. The costs will begin to fall (Figure 2L: red and brown lines) as the prevalence of infection falls. There will be an initial increase in costs as the backlog of untreated patients is taken up. The cumulative costs to 2025 would not increase. Angola currently spends about US$215 *per capita* on health so that a policy of early treatment will only increase overall costs by about 2% of total spending on health up to 2025 but after that there would be considerable cost savings.

## PMTCT

**Bringing it all together**

Figure 3 is a direct comparison of what would have happened under the *No ART* counterfactual, what has happened under the *Current Programme*, what should happen under the *Accelerated Programme*, and what could happen under the *Expanded Programme* with universal access and early treatment following the 2013 WHO guidelines.

The *Current Programme* has already averted many new infections and saved many lives, and will continue to save, both lives (Figure 3A and B) and money (Figure 3C and D) but the epidemic will continue indefinitely.

The *Expanded Programme* will incur a small increase in costs as treatment is expanded (red and green lines (Figure 3C and D) but will save many more lives (Figure 3A and B) and, in the long run, save more money (Figure 3C and D), eliminate HIV and bring an end to AIDS related deaths (Figure 3A and B).

## Discussion

Angola has a relatively small epidemic of HIV compared to other countries in southern Africa. The Angolan government has been successful in the extent to which it has made ART therapy available in the public sector. Currently, an estimated 42 thousand people are being kept alive on ART without increasing the cost. It is clear that expanding access will save more lives and, in the long run, will save money and eliminate HIV.

The Angolan Government is now committed to an *Accelerated Programme* of action which will double the number of people on ART by the end of 2015 but many logistical problems will have to be overcome. Two critical issues are drug supply and adherence. A regular and reliable supply of drugs must be assured. Stock-outs will create anger and mistrust among infected people and poor adherence, for any reason, will lead to viral rebound, treatment failure, on-going transmission and drug resistance. These two considerations must be at the forefront of plans to effectively control and eventually eliminate HIV. Given the shortage of trained clinicians the programme will have to depend heavily on community mobilization and the training and support of a cadre of community health workers. However this help to facilitate community involvement, empower local people and especially women, create jobs and stimulate local economies.

The *Accelerated Programme* will provide a sound basis for further expansion of treatment as described in the *Expanded Programme*. With currently available interventions universal access to early treatment is the only way to eliminate HIV in Angola and all HIV-positive people will die of AIDS related conditions if the do not start ART and starting ART as soon as possible after sero-conversion is in the best interests of the individual person. Other methods of support and control can and should play an important supporting role.[16] To achieve high levels of compliance it will be necessary to deal with problems of stigma and discrimination and to ensure that there is strong community involvement and support for people living with HIV.

While this paper focuses on treatment, continued scale-up of prevention interventions are critical to the HIV response. Condoms are highly effective if used properly and should be readily available to all that need them. Reducing the number of sexual partners and better control of other sexually transmitted infections are important in themselves and will contribute further to the control of HIV. Other prevention packages designed for key populations also need to be in place and can make an important contribution to stopping the epidemic of HIV; universal access to early treatment provides an ideal entry point for each of them. Finally, by developing programmes that are firmly based in local communities it will be possible to provide training and education while employing community outreach workers thereby creating jobs and stimulating local economies and ensuring sustainability. It is still important to confirm the validity or otherwise of these results and to improve the model predictions. Here we have used the reported coverage of ART but this has never been directly measured. To do so a sample of women who test positive for HIV each year should be tested for the presence of anti-retroviral drugs, their viral loads should be measured and an incidence assay should be used to estimate incidence. We have assumed a cost of US$250 per year for care and support of those infected with HIV. The results are sensitive to this assumption but this level of cost could be achieved if there is effective community support and mobilization. If this money were used to train and employ community health workers it would create jobs and provide an important stimulus to local economies in the poorest communities.

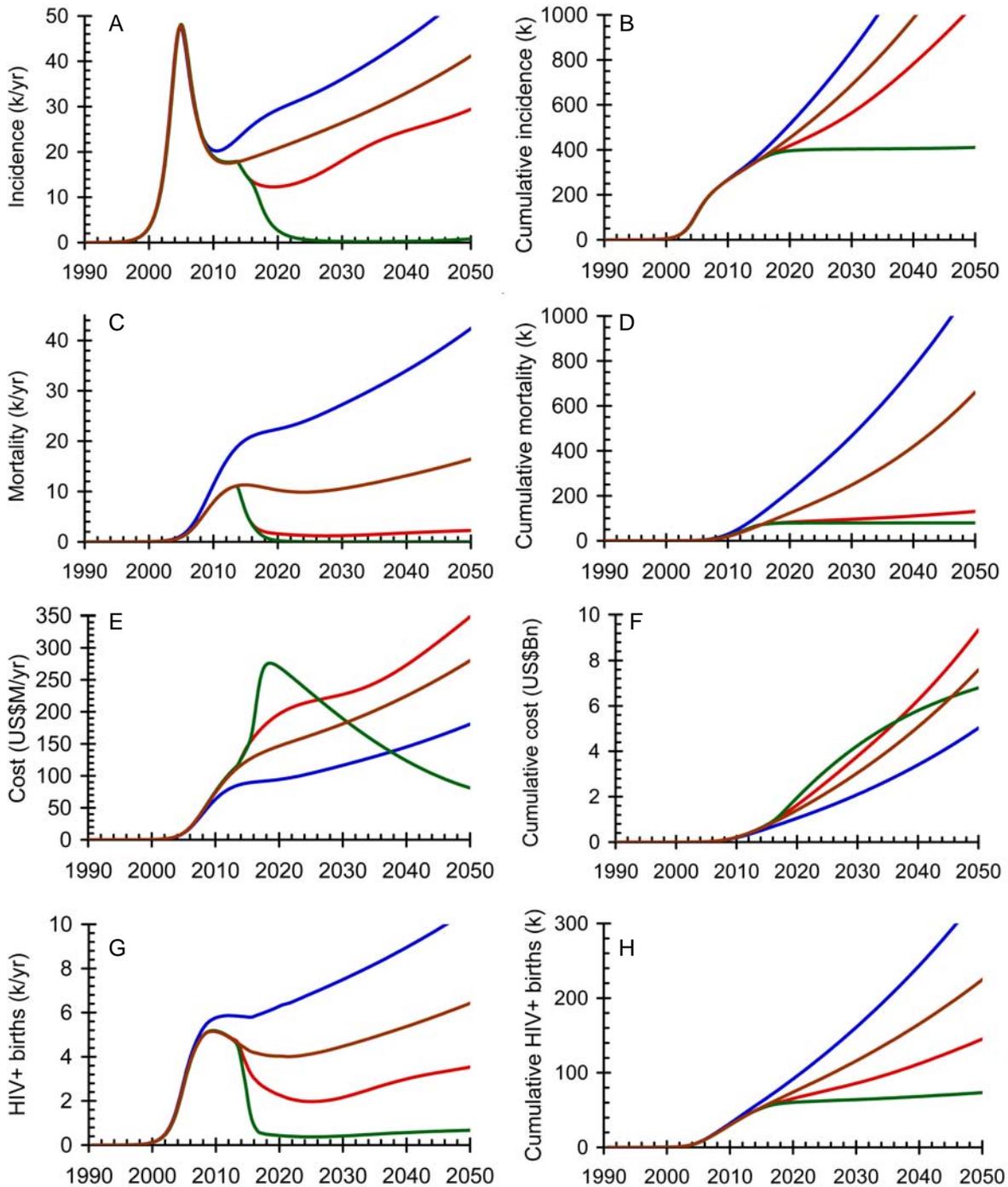

Figure 3. Left column annual rates; right column cumulative numbers. A & B: HIV incidence. C & D: AIDS deaths. E & F: Cost; G & H: HIV-positive births. Blue lines: *No ART* counterfactual. Brown line: *Current Programme*. Red line: *Accelerated Programme*. Green: *Expanded Programme*.

The Department of Health in Angola has already adopted the new WHO guidelines, with an opt-out approach for those with a CD4 cell count below 500 cells/µL. However, many facilities in Angola lack working CD4 machines. Since an estimated 90% of all HIV positive people will be eligible for immediate treatment according to the new WHO guidelines, it would be advisable to abandon the use of $CD4^+$ cell counts but to make viral load testing more widely available as a way of assessing the impact of the programme and of monitoring compliance. Under the *Expanded Programme* the rate at which people are started on ART will have to be increased by about ten times, but only for about two years as the intervention will rapidly reduce the incidence of new infections (Figure 2I). While this will require considerable organization it will only incur a marginal increase in costs for about five years and within a few years the savings in lives and money will be substantial.

What is needed now is informed leadership, an *Accelerated Programme* of ART provision, and if this can be successfully completed a further *Expanded Programme* of universal testing and immediate access to treatment, combined with good operational research, while monitoring the impact and implementation of the new programme. The sooner that this is started the more lives and money that will be saved. Angola has the wherewithal to stop the epidemic; all that is needed is the commitment and determination to achieve an AIDS Free Generation.

Table 5. Measured prevalence of HIV (%) among women attending ante-natal clinics at different facilities in the various provinces of Angola. Vertical grey lines indicate years in which no data were collected. Propn. gives the population in each province as a proportion of the total population.

| No. | Facilities | Province | Propn. | 1992 | 1993 | 1994 | 1995 | 1996 | 1998 | 1999 | 2001 | 2002 | 2004 | 2005 | 2007 | 2009 | 2011 | 2012 |
|---|---|---|---|---|---|---|---|---|---|---|---|---|---|---|---|---|---|---|
| 1 | Maternidade do Caxito | Bengo | 13.9 | | | | | | | | | | 1.20 | 1.81 | 4.51 | 3.25 | 3.00 | |
| 2 | Benguela | Benguela | 7.6 | | | | | | | | | | 0.60 | 2.80 | 2.41 | 5.00 | 4.60 | |
| 3 | Maternidade do Lobito | Benguela | | | | | | | | | | | 0.60 | 2.61 | 3.60 | 3.80 | 3.80 | |
| 4 | H.Prov.Kuito (Maternidade) | Bie | 4.5 | | | | | | | | | | 0.60 | 0.81 | 1.60 | 3.20 | 5.80 | |
| 5 | C.M.I. Cabinda | Cabinda | | | | | | | | | | 3.00 | 3.22 | 2.80 | 3.00 | 2.60 | 5.60 | |
| 6 | Cabinda Province | Cabinda | 2.8 | 6.00 | 7.00 | 7.00 | 5.00 | 8.00 | | 8.00 | | | | | | | | |
| 7 | Cacongo | Cabinda | | | | | | | | | | | | | 3.42 | 0.94 | 1.00 | 1.00 |
| 8 | Cahama | Cunene | | | | | | | | | | | | | 2.40 | 2.60 | 3.60 | 3.6 |
| 9 | H.Prov.Ondjiva(Maternidade) | Cunene | 2.6 | | | | | | | | | 12.00 | 8.80 | 10.62 | 9.40 | 7.40 | 8.22 | |
| 10 | Ombanja | Cunene | | | | | | | | | | | | | 4.00 | 3.17 | 2.41 | 2.41 |
| 11 | Maternidade do Huambo | Huambo | 6.5 | | | | | | | | | | 2.40 | 1.81 | 3.60 | 4.20 | 3.40 | |
| 12 | Matala | Huila | | | | | | | | | | | | | 1.20 | 1.60 | 2.40 | 2.4 |
| 13 | Maternidade do Lubango | Huila | 8.0 | | | | | | | | 4.00 | 1.00 | 2.60 | 4.03 | 3.01 | 2.80 | 4.39 | |
| 14 | H.Prov.Menongue(Maternidade) | Kuando Kubango | 1.6 | | | | | | | | | | 4.03 | 3.00 | 4.80 | 4.20 | 5.59 | |
| 15 | H.Central Ndalatando(Dr.A.A.Neto) | Kwanza Norte | 1.5 | | | | | | | | | | 1.00 | 1.20 | 1.60 | 1.00 | 2.00 | |
| 16 | Centro Policlínico da Cidade (Sumbe) | Kwanza Sul | 5.4 | | | | | | | | | | 0.80 | 1.40 | 1.40 | 1.00 | 1.40 | |
| 17 | Gabela | Kwanza Sul | | | | | | | | | | | | | 2.60 | 1.00 | 0.40 | 0.4 |
| 18 | C. S. Viana (Ana Paula) | Luanda | | | | | | | | | | | 2.20 | 2.81 | 1.80 | 3.00 | 4.78 | |
| 19 | C.S Ilha | Luanda | | | | | | | | | | | | | 2.86 | 3.40 | 5.20 | |
| 20 | C.S Samba | Luanda | | | | | | | | | | | | | 3.80 | 4.60 | 5.40 | |
| 21 | C.S. Asa Branca | Luanda | | | | | | | | | | 4.00 | 2.00 | 1.81 | 0.60 | 1.60 | 2.40 | |
| 22 | C.S. Cacuaco | Luanda | 22.2 | | | | | | | | | | 1.80 | 1.21 | 1.80 | 2.20 | 2.60 | |
| 23 | C.S. Hoji ya Henda | Luanda | | | | | | | | | | 2.00 | 3.80 | 4.41 | 1.80 | 3.20 | 1.41 | |
| 24 | H. Kilamba Kiaxi | Luanda | | | | | | | | | | 3.00 | 3.86 | 2.84 | 4.00 | 3.00 | 2.59 | |
| 25 | Luanda | Luanda | | | | | | | 1.00 | 3.00 | 3.00 | 8.00 | | | | | | |
| 26 | Lucrecia Paim | Luanda | | | | | | | | | | | 4.40 | 2.00 | | | | |
| 27 | Maternidade Ngangula | Luanda | | | | | | | | | | | 4.36 | 3.82 | | 4.80 | 2.81 | |
| 28 | H. Dundo (Maternidade) | Luanda Norte | 3.1 | | | | | | | | | | 3.40 | 3.41 | 6.60 | 5.60 | 6.36 | |
| 29 | Nzage-Cambulo | Luanda Norte | | | | | | | | | | | | | 3.60 | 2.87 | 3.80 | 3.8 |
| 30 | C.M.I. Saurimo | Lunda Sul | 1.5 | | | | | | | | | 1.00 | 3.42 | 3.60 | 4.60 | 2.80 | 4.00 | |
| 31 | Muconda | Lunda Sul | | | | | | | | | | | | | 0.75 | 2.23 | 2.99 | 2.99 |
| 32 | C.S. Ritondo | Malange | 2.9 | | | | | | | | | | 1.40 | 1.80 | 1.80 | 0.80 | 1.60 | |
| 33 | C.M.I. Luena | Moxico | 2.2 | | | | | | | | | | 2.61 | 2.00 | 3.20 | 3.00 | 3.01 | |
| 34 | Luau | Moxico | | | | | | | | | | | | | 2.40 | 0.60 | 2.99 | 2.99 |
| 35 | H. Mat.Infantil do Namibe | Namibe | 1.5 | | | | | | | | | | 2.00 | 3.60 | 2.12 | 3.70 | 4.01 | |
| 36 | Maternidade de Uige | Uige | 4.2 | | | | | | | | | | 4.90 | 1.00 | 0.60 | 1.00 | | |
| 37 | Negage | Uige | | | | | | | | | | | | | 1.80 | 1.00 | 0.60 | 0.6 |
| 38 | C.S 1º de Maio Zaire | Zaire | 1.6 | | | | | | | | | | 2.20 | 2.10 | 3.60 | 1.60 | 1.41 | |
| 39 | H.P Zaire | Zaire | | | | | | | | | | | | | 2.31 | 0.80 | 0.60 | |

## Appendix 1

HIV prevalence data for women attending ante-natal clinics are given in Table 5. Because the data are sparse we consider each data set individually. Full details of the analysis are available from the author on request. We first consider the timing of the epidemics in the different provinces. The only data sets in Table 5 for which we can identify the initial rise in the prevalence of HIV, with confidence, are those numbered 1, 2, 3, 6, 11, 25, 28, and 30. Fitting logistic curves to these data shows that in all but two facilities the mean year at which the epidemic reached half of its asymptotic value (*half-max.*) was 2004.1 (Range: 2002.6 to 2005.2; 95% confidence interval 2003.0 to 2005.2). There are two outliers: data set No. 6 from Cabinda and data set No. 25 from Luanda.

The epidemic of HIV in Cabinda appears to have reached *half-max.* in 1989.8, 14 years before the other provinces in Angola. The enclave of Cabinda is separated from the rest of Angola by a strip of territory belonging to the Democratic Republic of the Congo and it is likely that the epidemic started in Cabinda earlier than in the rest of Angola. A separate analysis is therefore needed for Cabinda and since Cabinda only makes up 1.8% of the total population, we exclude Cabinda from this analysis.

There are 10 data sets for Luanda but in only one of them, No. 25 in Table 5, can we establish the timing of the initial rise of the epidemic. This data set only contains four points and they suggest that the epidemic in Luanda reached *half-max.* in 1999, five years before the rest of the data sets. However this data set is in conflict with the other nine for Luanda as it suggests an asymptotic prevalence of 8.8% while the other nine suggest an asymptotic prevalence of 4.2% (95% CI: 2.8% to 5.6%). Since this data set contradicts the other nine data sets for Luanda as regards the asymptote and the other six data sets for the country as regards timing we exclude it from further analysis.* For data sets 1, 2, 3, 11, 28, and 30 in Table 5 we use the fitted values to determine the timing; for the remaining data sets, for which the timing cannot be determined, we use the average value estimated from these data sets and set *half-maximum.* to 2004.1. This gives the epidemic trend used in the analysis (Figure 2).

## Appendix 2

Analysing the data for Angola raises an important question. The initial rate of increase of the epidemic, from the best fit model has a doubling time of just less than one year which is about the same as it is in South Africa by way of comparison. However, the steady state prevalence in South Africa is about 15% and in Angola it is only 2.5%, about six times less. This suggests that only about one-sixth as many people (expressed as a proportion) are at risk of HIV in Angola as they are in South Africa. There are two standard ways of dealing with the problem of heterogeneity in the literature. The first assumes that there are two groups, one at a fixed risk and the other at no risk, the size of the risk group is then adjusted to match the observed peak prevalence. The second assumes that since those at high risk are likely to get infected earlier, the average population risk declines as the prevalence increases and the present author has generally assumed that transmission declines exponentially with prevalence. These two assumptions correspond to either a step-function distribution of risk or a risk that declines exponentially with prevalence. In the case of Angola the prevalence saturates at a very low level and, if we assume that transmission declines exponentially with prevalence, the model becomes unstable. We therefore assume that the decline in transmission with prevalence follows a Weibull survival curve with a variable parameter so that

$$\lambda(P) = e^{-\left(\frac{P}{\alpha}\right)^k}$$

where $\lambda$ is the transmission parameter and $P$ is the prevalence of HIV. The parameter $\alpha$ determines the level at which the prevalence peaks and the parameter $k$ determines the shape of the function. With $k = 1$ the function is exponential; with $k = \infty$ the function is a step function. The value of $k$ that gives the best fit to the data is 2.25.

## Appendix 3

The World Health Organization guidelines[15] of 2013 advise HIV-positive people to start ART if their $CD4^+$ cell count is less than $500/\mu L$, if they have TB or Hepatitis B, if they are pregnant, under the age of five years, or in a sero-discordant relationship. Data on the distribution of $CD4^+$ cell counts in HIV-negative people[19] suggest that about 80% of all those currently infected with HIV in South Africa and not on ART will have a $CD4^+$ cell count below $500/\mu L$. If we include the other groups of people that should start treatment irrespective of their $CD4^+$ cell count, then about 90% of all HIV positive people are currently eligible for ART.[4]

If there is a need to triage people in order to treat those at greatest risk first, the sensible way to do this would be on the basis of their viral load. Individual $CD4^+$ cell counts can vary by an order of magnitude within populations,[19] the mean $CD4^+$ cell count can vary by a factor of two between populations,[19] and survival is independent of the initial $CD4^+$ cell count.[19,20] $CD4^+$ cell counts therefore have very little prognostic value except in that unfortunate circumstance when the count is very low by which time an infected person is likely to be in WHO clinical stages III or IV and in need of immediate treatment anyway. Considerable savings in time, human resources and money could be had by abandoning the use of $CD4^+$ cell counts to decide on when to start ART. People with a high viral load, on the other hand, have a reduced life expectancy[21] and are more infectious than those with a low viral load.[22] If the availability of anti-retroviral drugs is limited it would therefore have the greatest benefit for individual patients and have the greatest impact on transmission if preference was given to people with high viral loads.[23]

## Appendix 4

The current cost of drugs in Angola is given in Table 6. The cost of ART drugs has dropped from US$10,000 *p.a.* in the year 2000 and has fallen to about US$100 *p.a.* in South Africa. The drugs costs in Table 6 are likely to fall considerably in the near future. For the purposes of this study we therefore include only the cost of first line drugs. Since, under option B+ for pregnant women by far the greatest costs will be incurred during the life-time spent on

---

\* Consultation with the Angolan Ministry of Health, the Instituto Nacional de Luta contra SIDA (INLS) and WHO confirmed this decision since most data in Angola were collected after 2004 and the early data from Luanda may include data from other provinces.

ART following the birth of the child. We therefore use the annual cost of first line drugs for all women who start on option B+

Table 6. Annual cost of anti-retroviral drugs, support which includes drug delivery and monitoring, and upstream costs (US$).

|  | Drugs | Support | Upstream | Total |
| --- | --- | --- | --- | --- |
| First line | 354 | 357 | 284 | 995 |
| Second line | 1059 | 350 | 564 | 1973 |
| Children | 163 | 300 | 185 | 648 |
| PMTCT | 764 | 350 | 446 | 1560 |

We combine 'support' and 'upstream' costs under the heading 'Community care and support' in Table 1.

We do not have estimates of the average number inpatient or outpatient days or primary health care visits for HIV-positive people who are or are not on ART and we use corresponding estimates for South Africa.

We estimate the annual costs per person reached or tested to be: community mobilization US$3.38; condom distribution and promotion US$0.18; testing and counselling US$14.87 for a total of US$18.43. In Table 1 we allocate US$20 for counselling and testing per person tested.

We estimate policy and program support expressed as a percentage over direct costs to be: creating an enabling environment 0.8%; program management 2.6%; research 0.7%; monitoring and evaluation 1.0%; strategic communication 2.3%; logistics 1.2% for a total of 8.6%. Since these costs only add an additional 8.6% to the total costs and given the great uncertainty in the cost estimates we do not include these costs in this calculation.